\documentclass[sigconf]{acmart}

  \renewcommand\footnotetextcopyrightpermission[1]{} 
\usepackage{balance}

\def\BibTeX{{\rm B\kern-.05em{\sc i\kern-.025em b}\kern-.08emT\kern-.1667em\lower.7ex\hbox{E}\kern-.125emX}}

 \usepackage{algorithm} 
\usepackage{algorithmic}
\usepackage{multirow}
\usepackage[english]{babel}
\usepackage{float}
\usepackage{breqn}
\usepackage{footnote}
\usepackage{subfig}
\usepackage{balance}
\usepackage[mathscr]{eucal}
\usepackage{blindtext, graphicx}
\usepackage{booktabs}
\usepackage{soul,color}
\usepackage{multirow}
\usepackage{algorithm,algorithmic}
\usepackage{bm}
\usepackage{ifpdf}
\usepackage{amsmath}
\usepackage{array}
\usepackage{multirow}
\usepackage{tikz}

\begin{document}

\title{Accelerating Bulk Bit-Wise X(N)OR Operation in Processing-in-DRAM Platform} 
\vspace{-0.7em}
 \author{Shaahin Angizi and Deliang Fan}
\affiliation{
	\institution{Department of Electrical and Computer Engineering, University of Central Florida, Orlando, FL 32816}}
\email{angizi@knights.ucf.edu, dfan@ucf.edu}

\renewcommand{\shortauthors}{Angizi and Fan}

\begin{abstract}

With Von-Neumann computing architectures struggling to address computationally- and memory-intensive big data analytic task today, Processing-in-Memory (PIM) platforms are gaining growing interests. In this way, processing-in-DRAM architecture has achieved remarkable success by dramatically reducing data transfer energy and latency. However, the performance of such system unavoidably diminishes when dealing with more complex applications seeking bulk bit-wise X(N)OR- or addition operations, despite utilizing maximum internal DRAM bandwidth and in-memory parallelism. 
In this paper, we develop \textit{DRIM} platform that harnesses DRAM as computational memory and transforms it into a fundamental processing unit. \textit{DRIM} uses the analog operation of DRAM sub-arrays and elevates it to implement bit-wise X(N)OR operations between operands stored in the same bit-line, based on a new dual-row activation mechanism with a modest change to peripheral circuits such sense amplifiers. The simulation results show that \textit{DRIM} achieves on average 71$\times$ and 8.4$\times$ higher throughput for performing bulk bit-wise X(N)OR-based operations compared with CPU and GPU, respectively. Besides, \textit{DRIM} outperforms recent processing-in-DRAM platforms with up to 3.7$\times$ better performance.
\end{abstract}
\maketitle
\vspace{-0.5em}

\section{Introduction}

In the last two decades, Processing-in-Memory (PIM) architecture, as a potentially viable way to solve the memory wall challenge, has been well explored for different applications \cite{chi2016prime,seshadri2017ambit,li2017drisa,angizi2017RIMPA,angizi2019mrima,angizi2018dima,angizi2018imce}. The key concept behind PIM is to realize logic computation within memory to process data by leveraging the inherent parallel computing mechanism and exploiting large internal memory bandwidth. The proposals for exploiting SRAM-based \cite{aga2017compute,eckert2018neural} PIM architectures can be found in recent literature. However, PIM in context of main memory (DRAM- \cite{li2017drisa,seshadri2017ambit,dai2018graphh}) has drawn much more attention in recent years mainly due to larger memory capacities and off-chip data transfer reduction as opposed to SRAM-based PIM. Such processing-in-DRAM platforms show significantly higher throughputs leveraging multi-row activation methods to perform bulk bit-wise operations by either modifying the DRAM cell and/or sense amplifier. For example, Ambit \cite{seshadri2017ambit} uses triple-row activation method to implement majority-based AND/OR logic, outperforming Intel Skylake-CPU, NVIDIA GeForce GPU, and even HMC \cite{HMC} by 44.9$\times$, 32.0$\times$, and 2.4$\times$, respectively. DRISA \cite{driskill2011latest} employs 3T1C- and 1T1C-based computing mechanisms and achieves 7.7$\times$ speedup and 15$\times$ better energy-efficiency over GPUs to accelerate convolutional neural networks.
However, there are different challenges in such platforms that make them inefficient acceleration solutions for X(N)OR- and addition-based applications such as DNA alignment and data encryption. Due to the intrinsic complexity of X(N)OR logic, current PIM designs are not able to offer a high-throughput X(N)OR-based operation despite utilizing the maximum internal bandwidth and memory level parallelism. This is because majority/AND/OR-based multi-cycle operations and required row initialization in the previous designs.

To overcome the memory bandwidth bottleneck and address the existing challenges, we propose a  high-throughput and energy-efficient PIM accelerator based on DRAM, called \textit{DRIM}. \textit{DRIM} exploits a new in-memory computing mechanism called Dual-Row Activation (DRA) to perform bulk bit-wise operations between operands stored in different word-lines. The DRA is developed based on analog operation of DRAM sub-arrays with a modest change in the sense amplifier circuit such that X(N)OR operation can be efficiently realized on every memory bit-line. In addition, such design addresses the reliability concerns regarding the voltage deviation on the bit-line and multi-cycle operations of the triple-row activation method.  We evaluate and compare \textit{DRIM}'s raw performance with conventional and PIM accelerators including a Core-i7 Intel CPU \cite{CPU}, an NVIDIA GTX 1080Ti Pascal GPU \cite{GPU1}, Ambit \cite{seshadri2017ambit}, DRISA-1T1C \cite{li2017drisa}, and HMC 2.0 \cite{HMC}, to handle bulk bit-wise operations. We observe that \textit{DRIM} achieves remarkable throughput compared to Von-Neumann computing systems (CPU/GPU) through unblocking the data movement bottleneck by on average  71$\times$/8.4$\times$ better throughput. \textit{DRIM} outperforms other PIMs in performing X(N)OR-based operations by up to 3.7$\times$ higher throughput. We further show that a 3D-stacked DRAM built on top of \textit{DRIM} can boost the throughput of the HMC by $\sim$13.5$\times$.  From the energy consumption perspective, \textit{DRIM} reduces the DRAM chip energy by 2.4$\times$ compared with Ambit \cite{seshadri2017ambit} and 69$\times$ compared with copying data through the DDR4 interface. 

To the best of our knowledge, this work is the first that designs a high-throughput and energy-efficient X(N)OR-friendly PIM architecture exploiting DRAM arrays. We develop \textit{DRIM} based on a set of novel microarchitectural and circuit-level schemes to realize a data-parallel computational unit for different applications.

\section{Background and Motivation} 
\subsection{Processing-in-DRAM Platforms} 

\begin{figure}[t]
\begin{center}
\begin{tabular}{c}
\includegraphics [width=0.98\linewidth]{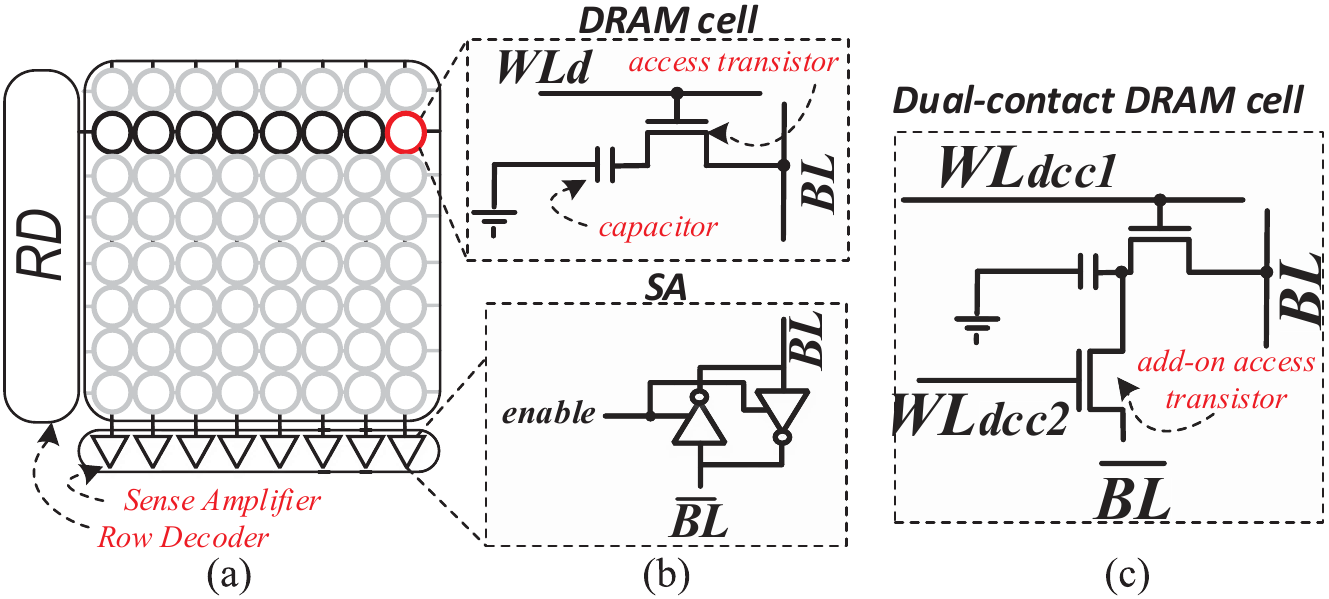}\vspace{-0.4em}
 \end{tabular} \vspace{-0.7em}
\caption{(a) DRAM sub-array organization, (b)  DRAM cell and Sense Amplifier, (c) Dual-contact DRAM cell.}\vspace{-1.2em}
\label{DRAM}
\end{center}
\end{figure}

A DRAM hierarchy at the top level is composed of channels, modules, and ranks. Each memory rank, with a data bus typically 64-bits wide, includes a set of memory chips that are manufactured with a variety of configurations and operate in unison \cite{kim2016ramulator,seshadri2017ambit}. Each chip is further divided into multiple memory banks that contains 2D sub-arrays of memory cells virtually-organized in memory matrices (mats). Banks within same chips share I/O, buffer and banks in different chips working in a lock-step manner. Each memory sub-array, as shown in Fig. \ref{DRAM}a, has 1) a large number of rows (typically $2^9$ or $2^{10}$) holding DRAM cells, 2) a row of Sense Amplifiers (SA), and 3) a Row Decoder (RD) connected to the cells. A DRAM cell basically consists of two elements, a capacitor (storage) and an Access Transistor (AT) (Fig. \ref{DRAM}b). The drain and gate of the AT is connected to the Bit-line ($BL$) and Word-line ($WL$), respectively. DRAM cell encodes the binary data by the charge of the capacitor. It represents logic `1' when the capacitor is full-charged, and logic `0' when there is no charge. 

$\bullet$\underline{\textbf{Write/Read Operation:}}
At initial state, both $BL$ and $\overline{BL}$ is always set to $\frac{V_{dd}}{2}$. Technically, accessing data from a DRAM's sub-array (write/read) after initial state is done through three consecutive commands \cite{seshadri2017ambit,seshadri2015fast} issued by the memory controller: 1) During the activation (i.e. {\tt ACTIVATE}), activating the target row, data is copied from the DRAM cells to the SA row. Fig. \ref{DRAM}b shows how a cell is connected to a SA via a $BL$. The selected cell (storing ${V_{dd}}$ or 0) shares its charge with the $BL$ leading to a small change in the initial voltage of $BL$ ($\frac{V_{dd}}{2}\pm \delta$). Then, by activating the $enable$ signal, the SA senses and amplifies the $\delta$ of the $BL$ voltage towards the original value of the data through voltage amplification according to the switching threshold of SA's inverter \cite{seshadri2015fast}. 2) Such data can be then transferred from/to SA to/from DRAM bus by a {\tt READ}/{\tt WRITE} command. In addition, multiple {\tt READ}/{\tt WRITE} commands can be issued to one row. 3) The {\tt PRECHARGE} command precharges both $BL$ and $\overline{BL}$ again and makes the sub-array ready for the next access.

$\bullet$\underline{\textbf{Copy and Initialization Operations:}}
To enable a fast ($<100ns$)  in-memory copy operation within DRAM sub-arrays, rather than using $\sim 1\mu s$ conventional operation in Von-Neumann computing systems, \textit{RowClone}-Fast Parallel Mode (FPM) \cite{seshadri2013rowclone} proposes a PIM-based mechanism that does not need to send the data to the processing units. In this scheme, two back-to-back {\tt ACTIVATE} commands to the source and destination rows without {\tt PRECHARGE} command in between, leads to a multi-kilo byte in-memory copy operation. This operation takes only $90ns$ \cite{seshadri2013rowclone}. This method has been further used for row initialization, where a preset DRAM row (either to `0' or `1') can be readily copied to a destination row.  RowClone imposes only a 0.01\% overhead to DRAM chip area \cite{seshadri2013rowclone}.

$\bullet$\underline{\textbf{Not Operation:}}
The {\tt NOT} function has been implemented in different works employing Dual-Contact Cells (DCC), as shown Fig. \ref{DRAM}c. DCC is mainly designed based on typical DRAM cell, but equipped with one more AT connected to $\overline{BL}$. Such hardware-friendly design \cite{seshadri2017ambit,kang2010one,lu2015improving} can be developed for a small number of rows on top of existing DRAM cells to enable efficient {\tt NOT} operation with issuing two back-to-back {\tt ACTIVATE} commands \cite{seshadri2017ambit}.  
In this way, the memory controller first activates the $WL_{dcc1}$ (Fig. \ref{DRAM}c) of input DRAM cell, and reads the data out to the SA through $BL$. It then activates $WL_{dcc2}$ to connect $\overline{BL}$ to the same capacitor and so writes the negated result back to the DCC.

\begin{figure}[t]
\begin{center}
\begin{tabular}{c}
\includegraphics [width=0.98\linewidth]{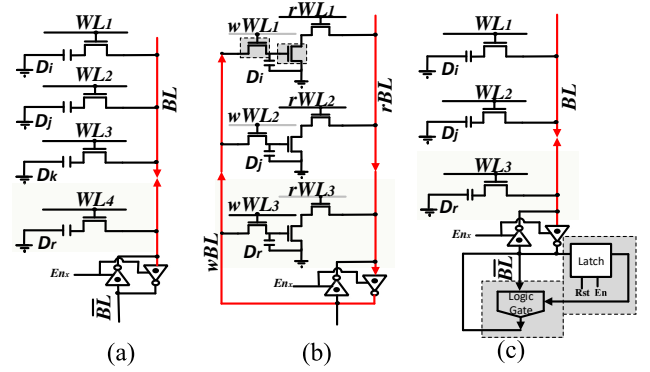}\vspace{-0.4em}
 \end{tabular} \vspace{-0.7em}
\caption{(a) Ambit's TRA \cite{seshadri2017ambit}, (b)  DRISA's 3T1C \cite{li2017drisa}, (c) DRISA's 1T1C-logic \cite{li2017drisa}. Glossary- $D_i$/$D_j$: input rows data, $D_k$: initialized row data, $D_r$ result row data.}\vspace{-2.3em}
\label{DRAM_back}
\end{center}
\end{figure}

$\bullet$\underline{\textbf{Other Logic Operations:}}
To realize the logic function in DRAM platform, \textit{Ambit} \cite{seshadri2017ambit} extends the idea of RowClone by implementing 3-input majority ({\tt Maj3})-based operations in memory by issuing the {\tt ACTIVATE} command to three rows simultaneously followed by a single {\tt PRECHARGE} command, so-called {\textit{Triple Row Activation (TRA)} method. As shown in Fig. \ref{DRAM_back}a, considering one row as the control, initialized by $D_k$= `0'/`1', Ambit can readily implement in-memory  {\tt AND2}/{\tt OR2} in addition to {\tt Maj3} functions through charge sharing between connected cells ($D_k$, $D_i$ and $D_j$) and write back the result on $D_r$ cell. It also leverage TRA mechanism along with DCCs to realize the complementary functions. However, despite Ambit shows only 1\% area over commodity DRAM chip \cite{seshadri2017ambit}, it suffers from multi-cycle PIM operations to implement other functions such as {\tt XOR2/XNOR2} based on TRA. Alternatively, \textit{DRISA}-3T1C method \cite{li2017drisa} utilizes the early 3-transistor DRAM design \cite{sideris1973intel}, in which the cell consists of two separated read/write ATs, and one more transistor to decouple the capacitor from the read $BL$ ($rBL$), as shown in Fig. \ref{DRAM_back}b. This transistor connects the two DRAM cells in a NOR style on the $rBL$ naturally performing functionally-complete {\tt NOR2} function. However, \textit{DRISA}-3T1C imposes very large area overhead (2T per cell) and still requires multi-cycle operations to implement more complex logic functions. \textit{DRISA}-1T1C method \cite{li2017drisa} offers to perform PIM through upgrading the SA unit by adding a CMOS logic gate in conjunction with a latch, as depicted in Fig. \ref{DRAM_back}c. Such inherently-multi-cycle operation can enhance the performance of a single function through add-on CMOS circuitry, in two consecutive cycles. In first cycle, $D_i$ is read out and stored in the latch, and in the second cycle, $D_j$ is sensed to perform the computation. However, this design imposes excessive cycles to implement other logic functions and at least 12 transistors to each SA. Recently, \textit{Dracc} \cite{deng2018dracc} implements a carry look-ahead adder by enhancing Ambit \cite{seshadri2017ambit} to accelerate convolutional neural networks.
\vspace{-0.7em}

\subsection{Challenges} 
There are three main challenges in the existing processing-in-DRAM platforms that make them inefficient acceleration solutions for XOR-based computations and we aim to resolve them:

$\bullet$\underline{\textbf{Limited throughput (Challenge-1):}} Due to the intrinsic complexity of X(N)OR-based logic implementations, current PIM designs (such as Ambit \cite{seshadri2017ambit}, DRISA \cite{li2017drisa}, and Dracc \cite{deng2018dracc}) are not able to offer a high-throughput and area-efficient X(N)OR or addition in-memory operation despite utilizing maximum internal DRAM bandwidth and memory-level parallelism for NOT, (N)AND, (N)OR, and MAJ/MIN logic functions. 
Moreover, while DRISA-1T1C method could implement either XNOR or XOR functions as the add-on logic gate, it requires at least two consecutive cycles to perform the computation, which in turn limits other logics implementation. We address this challenge by proposing the DRA mechanism in Section 3.1 and 3.4.

$\bullet$\underline{\textbf{Row initialization (Challenge-2):}} Given R=A$op$B  function ($op$ $\in$ {\tt AND2/OR2}}), TRA-based method \cite{seshadri2017ambit,seshadri2015fast} takes 4 consecutive steps to calculate one result as it relies on row initialization: 1-RowClone data of row A to row $D_i$ (Copying first operand to a computation row to avoid data-overwritten), 2-RowClone of row B to $D_j$, 3-RowClone of ctrl row to $D_k$ (Copying initialized control row to a computation row), 4-TRA and RowClone data of row $D_i$ to R row (Computation and Writing-back the result). Therefore TRA method needs averagely 360$ns$ to perform such in-memory operations. When it comes to {\tt XOR2/XNOR2} operation, Ambit requires at least three row-initialization steps to process two input rows. Obviously, this row-initialization load could adversely impact the PIM's energy-efficiency especially dealing with such big data problems. This challenge is addressed in Section 3.1 through the proposed sense amplifier, which totally eliminates the need for initialization in performing X(N)OR-based logics.

$\bullet$\underline{\textbf{Reliability concerns (Challenge-3):}} By simultaneously activating three cells in TRA method, the deviation on the $BL$ might be smaller than typical one-cell read operation in DRAM. This can elongate the sense amplification state or even adversely affect the reliability of the result \cite{seshadri2017ambit,seshadri2015fast}. The problem can be even intensified when multiple TRA are needed to implement X(N)OR-based computations. To explore and address this challenges, we perform an extensive Monte-Carlo simulation on our design in Section  3.3.

\vspace{-0.7em}

\section{DRIM Design}

\begin{figure}[t]
\begin{center}
\begin{tabular}{c}
\includegraphics [width=1\linewidth]{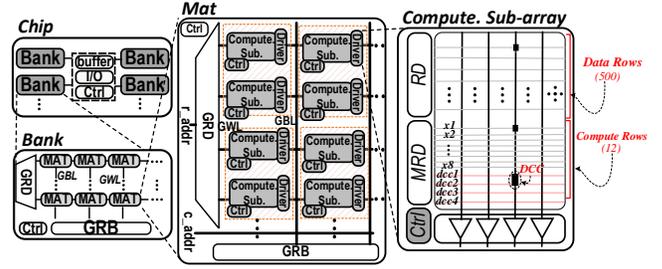}\vspace{-0.2em}
 \end{tabular} \vspace{-0.4em}
\caption{The \textit{DRIM} memory organization.}\vspace{-1.7em}
\label{arc}
\end{center}
\end{figure}

\textit{DRIM} is designed to be an independent, high- performance, energy-efficient accelerator based on main memory architecture to accelerate different applications. The main memory organization of \textit{DRIM} is shown in Fig. \ref{arc} based on typical DRAM hierarchy. Each mat consists of multiple computational memory sub-arrays connected to a Global Row Decoder (GRD) and a shared Global Row Buffer (GRB).   
According to the physical address of operands within memory, \textit{DRIM}'s Controller (Ctrl) is able to configure the sub-arrays to perform data-parallel intra-sub-array computations.
We divide the \textit{DRIM}'s sub-array row space into two distinct regions as depicted in Fig. \ref{arc}: 1- Data rows (500 rows out of 512) that include the typical DRAM cells (Fig. \ref{DRAM}b)  connected to a regular Row Decoder (RD), and 2- Computation rows (12), connected to a Modified Row Decoder (MRD), which enables multiple row activation required for bulk bit-wise in-memory operations between operands. Eight computational rows ($x1,...,x8$) include typical DRAM cells and four rows ($dcc1,...,dcc4$) are allocated to DCCs (Fig. \ref{DRAM}c) enabling {\tt NOT} function in every sub-array. \textit{DRIM}'s computational sub-array is motivated by Ambit \cite{seshadri2017ambit}, but enhanced and optimized to perform both TRA and the proposed \textit{Dual-Row Activation} (DRA) mechanisms leveraging charge-sharing among different rows to perform logic operations, as discussed below.  \vspace{-0.65em}

\subsection{New In-Memory Operations}
 
\begin{figure}[b]
\begin{center}
\begin{tabular}{c}
\includegraphics [width=0.99\linewidth]{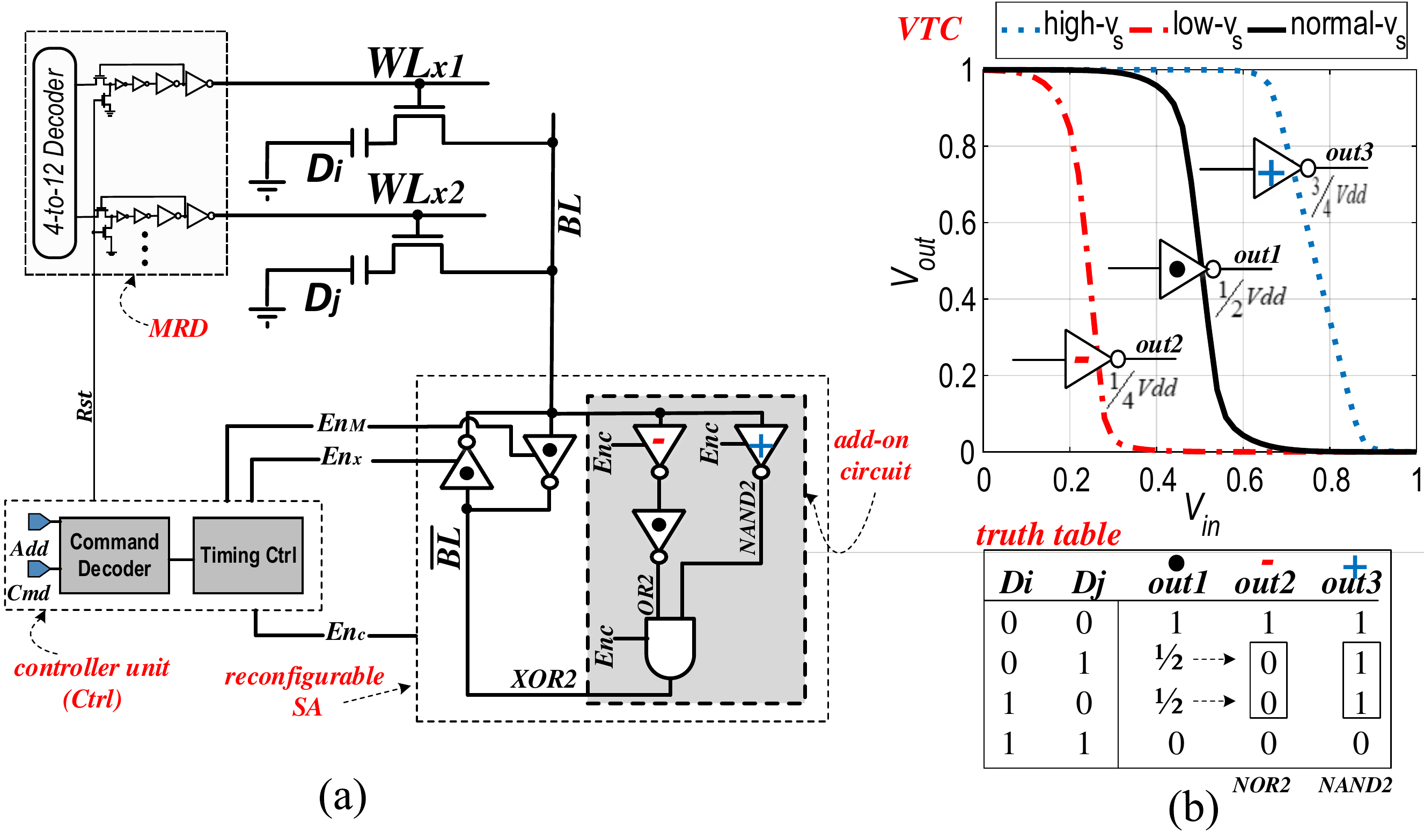}\\
 \end{tabular}\vspace{-0.8em}
\caption{(a) New sense amplifier design for \textit{DRIM}, (b) VTC and truth table of the SA's inverters.}
\label{NEWSA}
\end{center}\vspace{-1.2em}
\end{figure}

$\bullet$\underline{\textbf{Dual-Row Single-Cycle In-Memory X(N)OR:}}
With a careful observation on the existing processing-in-DRAM platforms, we realized that they are not able to efficiently handle two main functions prerequisite for accelerating a variety of applications (XNOR, addition). As a result, such platforms impose an excessive latency and energy to memory chip, which could be alleviated by rethinking about SA circuit. Our key idea is to perform in-memory {\tt XNOR2} through a DRA method to alleviate and address three of the challenges discussed in Section 2.3. To achieve this goal, we propose a new reconfigurable SA, as shown in Fig. \ref{NEWSA}a, developed on top of the existing DRAM circuitry. It consists of a regular DRAM SA equipped with add-on circuits including three inverters and one AND gate controlled with three enable signals ($En_M$,$En_x$,$En_C$). This design leverages the charge-sharing feature of DRAM cell and elevates it to implement {\tt XNOR2} logic between two selected rows through static capacitive-NAND/NOR functions in a single cycle. To implement capacitor-based logics, we use two different inverters with shifted Voltage Transfer Characteristic (VTC), as shown in Fig. \ref{NEWSA}b.

In this way, a NAND/NOR logic can be readily carried out based on high switching voltage ($V_s$)/low-$V_s$ inverters with standard high-$V_{th}$/low-$V_{th}$ NMOS and low-$V_{th}$/high-$V_{th}$ PMOS transistors. It is worth mentioning that, utilizing low/high-threshold voltage transistors a long with normal-threshold transistors have been accomplished in low-power application, and many circuits have enjoyed this technique in low-power design \cite{allam2000high,mutoh19951,kuroda19960,navi2009novel}.

\begin{figure}[h]
\begin{center}
\begin{tabular}{c}
\includegraphics [width=1\linewidth]{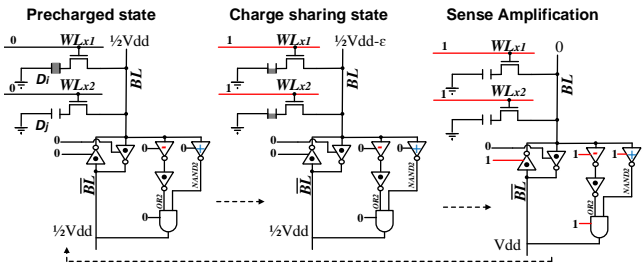}\\
 \end{tabular}\vspace{-0.8em}
\caption{Dual Row Activation to realize {\tt XNOR2}.}
\label{AND}
\end{center}\vspace{-1.2em}
\end{figure}

Consider $D_i$ and $D_j$ operands are RowCloned from data rows to $x1$ and $x2$ rows and both $BL$ and $\overline{BL}$ are precharged to $\frac{V_{dd}}{2}$ (Precharged State in Fig. \ref{AND}). To implement DRA, \textit{DRIM}'s ctrl first activates two $WL$s in computational row space (here, $x1$ and $x2$) through the modified decoder for charge-sharing when all the other enable signals are deactivated (Charge Sharing State). 
During Sense Amplification State, by activating the corresponding enable signals ($En_C$ and $En_x$) tabulated in Table \ref{ctrl}, the input voltage of both low- and high-$V_s$ inverters in the reconfigurable SA can be simply derived as  $V_i=\frac{n.V_{dd}}{C}$, where $n$ is the number of DRAM cells storing logic `1' and $C$ represents the total number of unit capacitors connected to the inverters (i.e. 2 in DRA method).

\begin{table}[h]
\caption{Control bits status in Sense Amplification state.}\vspace{-0.6em}
\centering
\scalebox{0.73}{
\begin{tabular}{|c|c|c|c|}
\hline
In-memory operations & $EN_M$ & $EN_x$ & $EN_C$ \\ \hline
W/R - Copy - NOT - TRA      & 1      & 1      & 0      \\ \hline
DRA                  & 0      & 1      & 1      \\ \hline
\end{tabular}}\vspace{-1em}
\label{ctrl}
\end{table}
Now, the low-$V_{s}$ inverter acts as a threshold detector by amplifying deviation from $\frac{1}{4}V_{dd}$ and realizes a {\tt NOR2} function as tabulated in the truth table in Fig. \ref{NEWSA}b. At the same time the high-$V_{s}$ inverter amplifies the deviation from $\frac{3}{4}V_{dd}$ and realizes a {\tt NAND2} function. Accordingly, {\tt XOR2} and {\tt XNOR2} functions of input operands can be realized after CMOS AND gate, respectively, on the $\overline{BL}$ and ${BL}$ based on Equation-(1) in a single memory cycle.  \vspace{-1.5em}

\begin{equation}
\begin{split}
  \footnotesize  \overline{BL}=(\overline{D_i.D_j}).(D_i+D_j)=D_i.\overline{D_j}+D_j.\overline{D_i}=D_i \oplus D_j\\
 \footnotesize  \Rightarrow  BL= D_i \odot D_j
    \end{split}
\end{equation}

\textit{DRIM}'s reconfigurable SA is especially optimized to accelerate {\tt X(N)OR2} operations, as well as supporting other memory and in-memory operations (i.e. Write/Read, Copy, NOT, and TRA). \textit{DRIM} ctrl activates $En_M$ and $En_x$ control-bits simultaneously (when $En_C$ is deactivated) to perform such operations. However, in this work, we only use Ambit's TRA mechanism to directly realize in-memory majority function ({\tt Maj3}).

\begin{figure}[t]
\begin{center}
\begin{tabular}{c}
\includegraphics [width=0.98\linewidth]{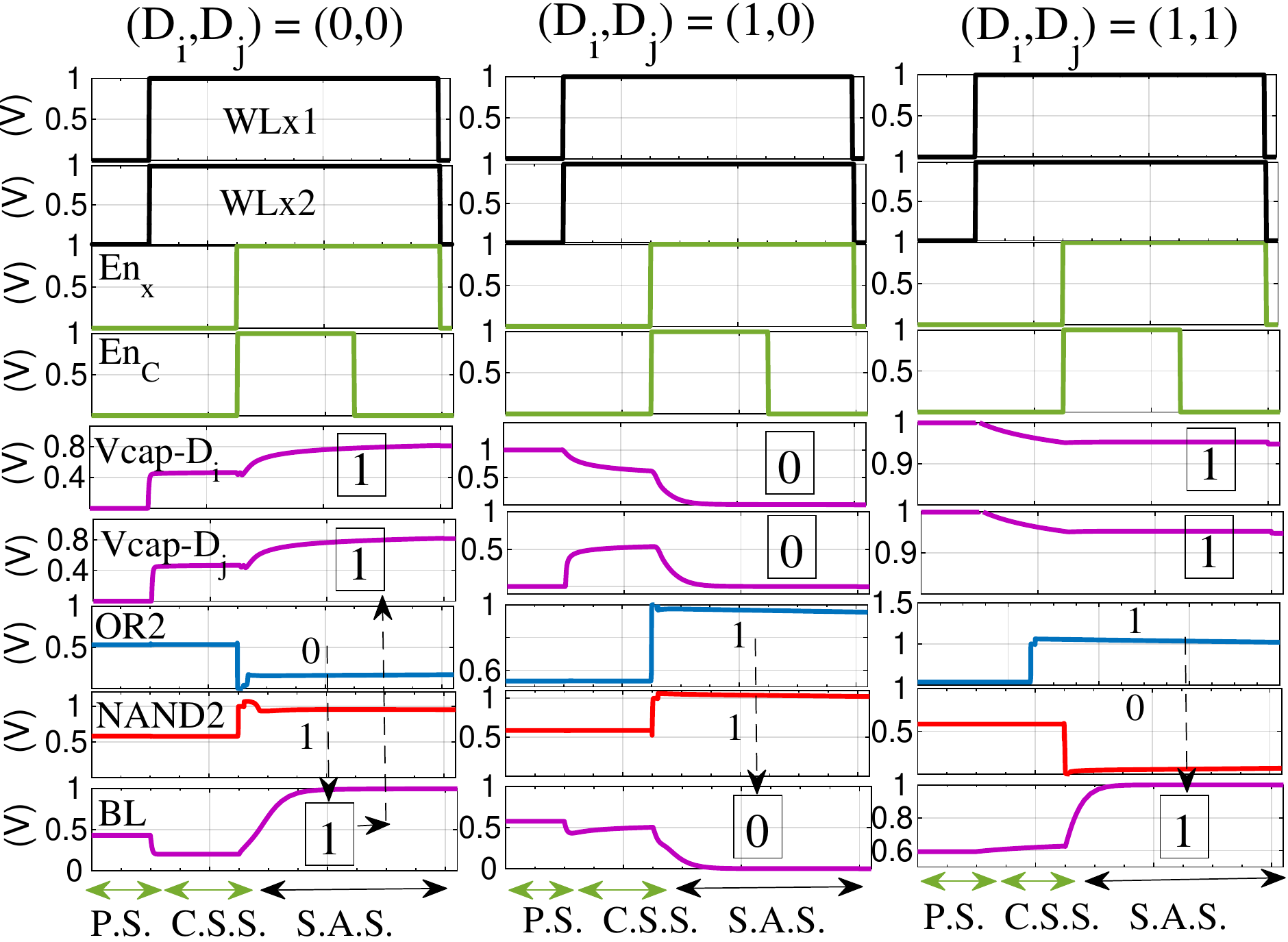}\\
 \end{tabular}\vspace{-0.8em}
\caption{The transient simulation of the internal \textit{DRIM}'s sub-array signals involved in DRA mechanism. Glossary- Vcap-$D_i$ and Vcap-$D_J$ represent the voltage across the two selected DRAM cell's capacitors connected to WLx1 and WLx2. P.S., C.S.S., S.A.S. are short for Precharged State, Charge Sharing State, and Sense Amplification State, respectively.}
\label{tran}
\end{center}\vspace{-1.2em}
\end{figure}

The transient simulation results of DRA method to realize single-cycle in-memory {\tt XNOR2} operation is shown in Fig. \ref{tran}. We can observe how $BL$ voltage and accordingly cell's capacitor is charged to $V_{dd}$ (when $D_iD_j$=00/11) or discharged to GND (when $D_iD_j$=01/10) during sense amplification state. Therefore, DRA method can effectively provide a single-cycle X(N)OR logic to address the challenge-1 and -2 discussed in Section 2.3 by eliminating the need for multiple TRA- \cite{seshadri2017ambit} or NOR-based \cite{li2017drisa} operations as well as row initialization steps.

$\bullet$\underline{\textbf{In-Memory Adder:}}
\textit{DRIM}'s sub-array can perform addition /subtraction ({\tt add/sub}) operation quite efficiently. Assume $D_i,D_j$, and $D_k$ as input operands, the carry-out ($C_{out}$) of the Full-Adder (FA) can be directly generated through $MAJ3(D_i,D_j,D_k)=D_iD_j+D_iD_k+D_jD_k$ using TRA method. Moreover, the $Sum$ can be readily carried out through two back-to-back {\tt XOR2} operations based on the proposed DRA mechanism.

\subsection{ISA Support}

While \textit{DRIM} is meant to be an independent high-performance and energy-efficient accelerator, we need to expose it to programmers and system-level libraries to utilize it. From a programmer perspective, \textit{DRIM} is more of a third party accelerator that can be connected directly to the memory bus or through PCI-Express lanes rather than a memory unit, thus it is integrated similar to that of GPUs. Therefore, a virtual machine and ISA for general-purpose parallel thread execution need to be defined similar to PTX \cite{GPU} for NVIDIA. Accordingly, the programs are translated at install time to the \textit{DRIM} hardware instruction set discussed here to realize the functions tabulated in Table \ref{ISA}. The micro and control transfer instructions are not discussed here.

\begin{table}[b]
\caption{The basic functions supported by \textit{DRIM}.}
\centering
\scalebox{0.66}{
\begin{tabular}{|c|c|c|c|}
\hline
Func.               & Operation                                                                                                                   & Command Sequence                                                                                                                                                                                                               & {\tt AAP} Type                                                    \\ \hline
copy                & $D_r \leftarrow D_i$                                                                                                        & {\tt AAP}($D_i,D_r$)                                                                                                                                                                                                           & 1                                                                 \\ \hline
NOT                 & $D_r \leftarrow \overline{D_i}$                                                                                             & \begin{tabular}[c]{@{}c@{}}{\tt AAP}($D_i,dcc2$)\\ {\tt AAP}($dcc1,D_r$)\end{tabular}                                                                                                                                          & \begin{tabular}[c]{@{}c@{}}1\\ 1\end{tabular}                     \\ \hline
MAJ/MIN$\dagger$    & $D_r\leftarrow MAJ3(D_i, D_j,D_k)$                                                                                          & \begin{tabular}[c]{@{}c@{}}{\tt AAP}($D_i,x1$)\\ {\tt AAP}($D_j,x2$)\\ {\tt AAP}($D_k,x3$)\\ {\tt AAP}($x1,x2,x3,D_r$)\end{tabular}                                                                                            & \begin{tabular}[c]{@{}c@{}}1\\ 1\\ 1\\ 4\end{tabular}             \\ \hline
XNOR2/XOR2$\dagger$ & $D_r \leftarrow D_i \odot D_j$                                                                                              & \begin{tabular}[c]{@{}c@{}}{\tt AAP}($D_i,x1$)\\ {\tt AAP}($D_j,x2$)\\ {\tt AAP}($x1,x2,D_r$)\end{tabular}                                                                                                                     & \begin{tabular}[c]{@{}c@{}}1\\ 1\\ 3\end{tabular}                 \\ \hline
Add/Sub$\dagger$             & \begin{tabular}[c]{@{}c@{}}$Sum \leftarrow D_i \oplus D_j \oplus D_k$\\ $C_{out}\leftarrow MAJ3(D_i, D_j,D_k)$\end{tabular} & \begin{tabular}[c]{@{}c@{}}{\tt AAP}($D_i,x1,x2$)\\ {\tt AAP}($D_j,x3,x4$)\\ {\tt AAP}($D_k,x5,x6$)\\ {\tt AAP}($x2,x4,dcc2$)\\ {\tt AAP}($x6,dcc1,dcc4$)\\ {\tt AAP}($dcc3,Sum$)\\ {\tt AAP}($x1,x2,x3,C_{out}$)\end{tabular} & \begin{tabular}[c]{@{}c@{}}2\\ 2\\ 2\\ 3\\ 3\\ 1\\ 4\end{tabular} \\ \hline
\end{tabular}
}
\label{ISA}

$\dagger$ \scriptsize{Complement functions and Subtraction can be realized with $dcc$ rows.}\vspace{-1.5em}
\end{table}

\textit{DRIM} is developed based on {\tt ACTIVATE-ACTIVATE-PRECHARGE} command a.k.a. {\tt AAP} primitives and most bulk bitwise operations involve a sequence of {\tt AAP} commands. To enable processor to efficiently communicate with \textit{DRIM}, we developed four types of {\tt AAP}-based instructions that only differ from the number of activated source or destination rows: 

1- {\tt AAP (src, des, size)} that runs the following commands sequence: 1) {\tt ACTIVATE} a source address ({\tt src}); 2) {\tt ACTIVATE} a destination address ({\tt des}); 3) {\tt PRECHARGE} to prepare the array for the next access. The {\tt size} of input vectors for in-memory computation must be a multiple of DRAM row size, otherwise the application must pad it with dummy data. The type-1 instruction is mainly used for copy and NOT functions; 
2- {\tt AAP (src, des1, des2, size)}, 1) {\tt ACTIVATE} a source address; 2) {\tt ACTIVATE} two destination addresses; 3) {\tt PRECHARGE}. This instruction copies a source row simultaneously to two destination rows; 
3- {\tt AAP (src1, src2, des, size)} that performs DRA method by activating two source addresses and then writes back the result on a destination address;
4- {\tt AAP (src1, src2, src3, des, size)} that performs Ambit-TRA method \cite{seshadri2017ambit} by activating three source rows and writing back the {\tt MAJ3} result on a destination address.

For instance, in order to implement the addition-in-memory, as shown in Table \ref{ISA}, three {\tt AAP}-type2 commands double-copy the three input data rows to computational rows ($x1,..,x6$). Then, the $Sum$ function is realized through two back-to-back {\tt XOR2} operations with {\tt AAP}-type3. The $C_{out}$ is generated by {\tt AAP}-type4 and written back to the designated data row. \vspace{-0.5em}

\subsection{Reliability}

We performed a comprehensive circuit-level simulation to study the effect of process variation on both DRA and TRA methods considering different noise sources and variation in all components including DRAM cell ($BL$/$WL$ capacitance and transistor, shown in Fig. \ref{cap}) and SA (width/length of transistors-$V_s$). We ran Monte-Carlo simulation in Cadence Spectre with 45nm NCSU Product Development Kit (PDK) library \cite{NCSU_PDK} (DRAM cell parameters were taken and scaled from Rambus \cite{Rambus}) under 10000 trials and increased the amount of variation from $\pm$0\% to $\pm$30\% for each method. Table \ref{var} shows the percentage of the test error in each variation. We observe that even considering a significant $\pm$10\%  variation, the percentage of erroneous DRA across 10000 trials is 0\%, where TRA method shows a failure with 0.18\%. 
Therefore, \textit{DRIM} offers a solution to alleviate challenge-3 by showing an acceptable voltage margin in performing operations based on DRA mechanism. By scaling down the transistor size, the process variation effect is expected to get worse \cite{seshadri2013rowclone,seshadri2017ambit}. Since \textit{DRIM} is mainly developed based on existing DRAM structure and operation with slight modifications, different methods currently-used to tackle process variation can be also applied for \textit{DRIM}. Besides, just like Ambit, \textit{DRIM} chips that fail testing due to DRA or TRA methods can be potentially considered as regular DRAM chips alleviating DRAM yield. 


\begin{table}[h]
\begin{minipage}[b]{0.5\linewidth}
\centering
\includegraphics[width=0.8\linewidth]{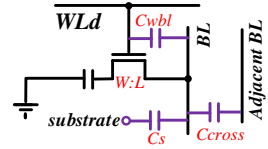}
\vspace{-0.5em}
    \captionof{figure}{Noise sources in DRAM cell. Glossary: $Cwbl$, $Cs$, and $Ccross$ are $WL$-$BL$, $BL$-substrate, and $BL$-$BL$ capacitance, respectively.}
    \label{cap}
\end{minipage}
\begin{minipage}[b]{0.43\linewidth}
\centering
    \small\vspace{-1em}
\begin{tabular}{|c|c|c|}
\hline
Variation & TRA  & DRA  \\ \hline
$\pm$5\%  & 0.00 & 0.00 \\ \hline
$\pm$10\% & 0.18 & 0.00 \\ \hline
$\pm$15\% & 5.5  & 1.2  \\ \hline
$\pm$20\% & 17.1 & 9.6  \\ \hline
$\pm$30\% & 28.4 & 16.4  \\ \hline
\end{tabular}\vspace{3em}
    \caption{Process variation analysis.}\vspace{-1em}
    \label{var}
\end{minipage}\hfill
\end{table}

\subsection{Performance}

\begin{figure}[h]
\begin{center}
\begin{tabular}{c}
\includegraphics [width=0.95\linewidth,  height=3.6cm]{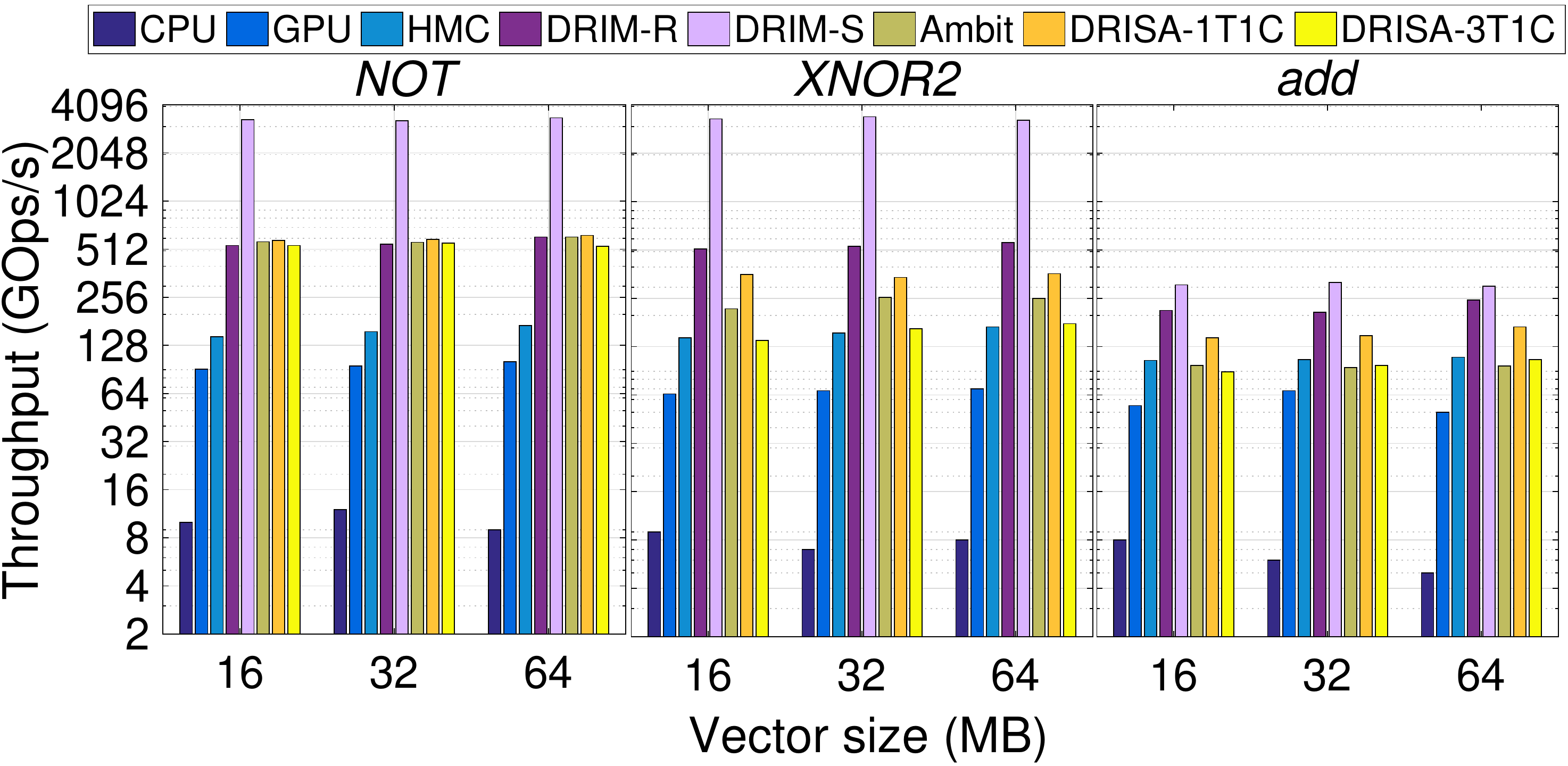}\\
 \end{tabular}\vspace{-0.8em}
\caption{Throughput of different platforms (Y: log scale).}
\label{measure}
\end{center}\vspace{-1.5em}
\end{figure}

$\bullet$\underline{\textbf{Throughput:}} We evaluate and compare the \textit{DRIM}'s raw performance with conventional computing units including a Core-i7 Intel CPU \cite{CPU} and an NVIDIA GTX 1080Ti Pascal GPU \cite{GPU1}.
There is a great deal of PIM accelerators that present reconfigurable platforms or application-specific logics in or close to memory die \cite{angizi2018design,bojnordi2016memristive,angizi2018cmp,ahn2016scalable,ahn2015pim,akin2015data,balasubramonian2014near,boroumand2017lazypim,farmahini2015nda,guo2015enabling,hsieh2016accelerating,kim2016neurocube,nair2015active,pattnaik2016scheduling,pugsley2014comparing,trancoso2015moving,tang2017data,zhang2014top,akerib2015non,angizi2018pima,parveen2018imcs2}. Due to the lack of space, we shall restrict our comparison to four recent processing-in-DRAM platforms, Ambit \cite{seshadri2017ambit}, DRISA-1T1C \cite{li2017drisa}, DRISA-3T1C \cite{li2017drisa}, and HMC 2.0 \cite{HMC}, to handle three main bulk bit-wise operations, i.e. {\tt NOT}, {\tt XNOR2}, and {\tt add}. To have a fair comparison, we report \textit{DRIM}'s and other PIM platforms' raw throughput implemented with 8 banks with 512$\times$256 computational sub-arrays. We further develop a 3D-Stacked DRAM with 256 banks in 4GB capacity similar to that of HMC 2.0 for the \textit{DRIM} (i.e. \textit{DRIM}-S) considering its computational capability. The Intel CPU consists of 4 cores and 8 threads working with two 64-bit DDR4-1866/2133 channels. The Pascal GPU has 3584 CUDA cores running at 1.5GHz \cite{GPU1} and 352-bit GDDR5X. The HMC has 32 vaults each with 10 GB/s bandwidth. Accordingly, we develop an in-house benchmark to run the operations repeatedly for $2^{27}$/$2^{28}$/$2^{29}$-length input vectors and report the throughput of each platform, as shown in Fig. \ref{measure}.\vspace{-0.7em}

We observe that  1) either the external or internal DRAM bandwidth has limited the throughput of the CPU, GPU, and even HMC platforms. However, HMC outperforms the CPU and GPU with $\sim$25$\times$ and 6.5$\times$ higher performance on average for bulk bit-wise operations. Besides, PIM platforms achieve remarkable throughput compared to Von-Neumann computing systems (CPU/GPU) through unblocking the data movement bottleneck. Regular \textit{DRIM} (\textit{DRIM}-R) shows on average  71$\times$ and 8.4$\times$ better throughput compared to CPU and GPU, respectively.
2) while \textit{DRIM}-R, Ambit, and DRISA platforms achieve almost the same performance on performing bulk bit-wise {\tt NOT} function, \textit{DRIM}-R outperforms other PIMs in performing {\tt X(N)OR2}-based operations. Our platform improves the throughput 2.3$\times$, 1.9$\times$, 3.7$\times$ compared with Ambit \cite{seshadri2017ambit}, DRISA-1T1C \cite{li2017drisa}, and DRISA-3T1C \cite{li2017drisa}, respectively.
3) \textit{DRIM}-S can boost the throughput of the HMC by 13.5$\times$. To sum it up, \textit{DRIM}'s DRA mechanism could effectively address challenge-1 by proposing the high-through bulk bit-wise X(N)OR-based operation. 

\begin{figure}[h]
\begin{center}
\begin{tabular}{c}
\includegraphics [width=0.6\linewidth,  height=3.3cm]{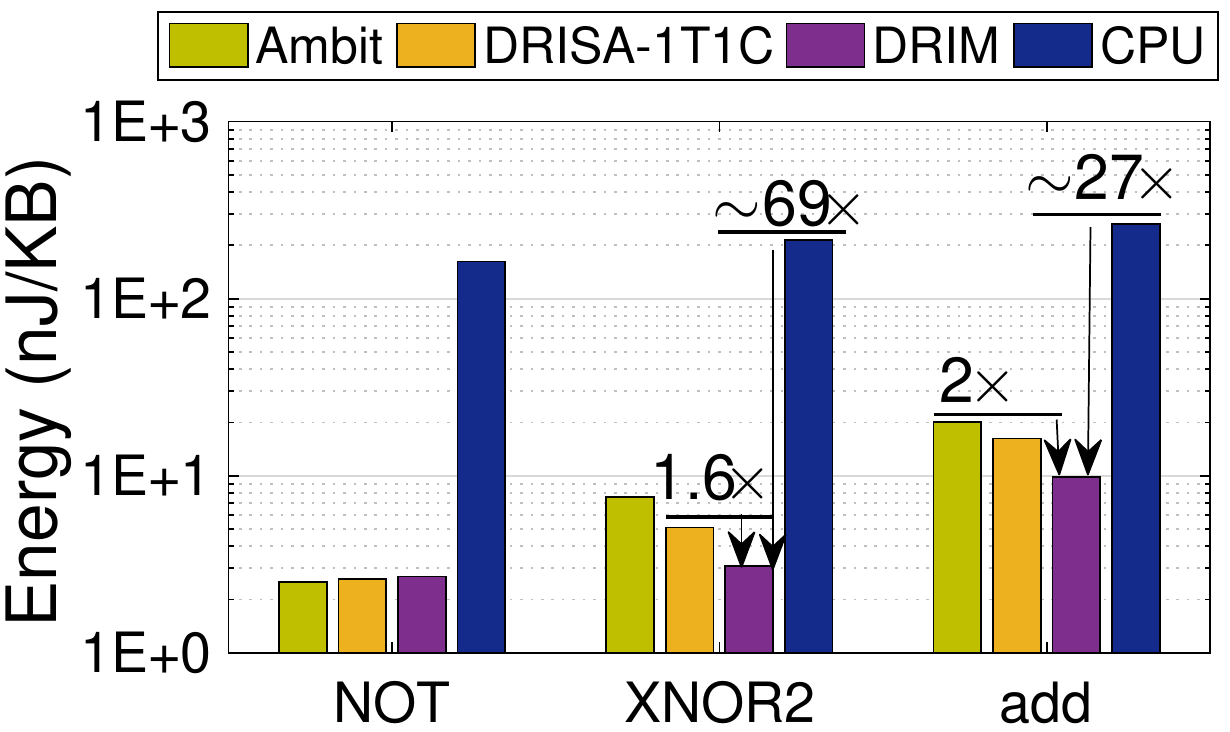}\\
 \end{tabular}\vspace{-0.8em}
\caption{Energy consumption of different platforms (Y: log scale).}
\label{energy}
\end{center}\vspace{-1.2em}
\end{figure}

$\bullet$\underline{\textbf{Energy:}}
We estimate the energy that DRAM chip consumes to perform the three bulk bit-wise operations per Kilo-Byte for \textit{DRIM}, Ambit \cite{seshadri2017ambit}, DRISA-1T1C \cite{driskill2011latest}, and CPU\footnote{This energy doesn't involve the energy that processor consumes to perform the operation.}. Note that, other operations such {\tt AND2/NAND2} and {\tt OR2/NOR2} in \textit{DRIM} can be built on top of TRA method with almost the same energy consumption to that of Ambit. 
Fig. \ref{energy} shows that \textit{DRIM} achieves  2.4$\times$ and  1.6$\times$ energy reduction over Ambit \cite{seshadri2017ambit} and  DRISA-1T1C \cite{driskill2011latest}, respectively, to perform bulk bit-wise {\tt XNOR2} operation. Besides, compared with copying data through the DDR4 interface, \textit{DRIM} reduces the energy by 69$\times$. As for bit-wise in-memory {\tt add} operation, \textit{DRIM} outperforms Ambit,  DRISA-1T1C, and CPU, respectively, with $\sim$2$\times$, 1.7$\times$, and 27$\times$ reduction in energy consumption. \vspace{-1em}

$\bullet$\underline{\textbf{Area:}}
To assess the area overhead of  \textit{DRIM} on top of commodity DRAM chip, four hardware cost sources must be taken into consideration. First, add-on transistors to SAs; in our design, each SA requires 22 additional transistors connected to each $BL$. Second, two rows of DCCs with two $WL$ associated with each; based on the estimation made by \cite{kang2010one}, each DCC row imposes roughly one transistor over regular DRAM cell to each $BL$. Third, the 4:12 MRD overhead (originally 4:16); we modify each $WL$ driver by adding two more transistors in the typical buffer chain, as depicted in Fig. \ref{NEWSA}a. Fourth, the Ctrl's overhead to control enable bits; ctrl generates the activation bits with MUX units with 6 transistors. To sum it up, \textit{DRIM} roughly imposes 24 DRAM rows per sub-array, which can be interpreted as $\sim 9.3\%$ of DRAM chip area.

\section{Discussion}

$\bullet$\underline{\textbf{Virtual Memory:}}
\textit{DRIM} has its own ISA with operations that can potentially use virtual addresses. To use virtual addresses, \textit{DRIM}'s ctrl must have the ability to translate virtual addresses to physical addresses. While in theory this looks as simple as passing the address of the page table root to \textit{DRIM} and giving \textit{DRIM}'s ctrl the ability to walk the page table, it is way more complicated in real-world designs. The main challenge here is that the page table can be scattered across different DIMMs and channels, while \textit{DRIM} operates within a memory module. Furthermore, page table coherence issues can arise. The other way to implement translation capabilities for \textit{DRIM} is through memory controller pre-processing of instructions being written to \textit{DRIM} instruction registers. For instance, if the programmer writes instruction {\tt APP (src,dec,256)}, then the memory controller intercepts the virtual addresses and translates them into physical addresses. Note that most systems have near memory controller translation capabilities, mainly to manage IOMMU and DMA accesses from I/O devices. One issue that can arise is that some operations are appropriate only if the resulting physical addresses are within specific plane, e.g., within the same bank. Accordingly, the compiler and the OS should work together to ensure that the operands of commands will result physical addresses that are suitable to the operation type. 


$\bullet$\underline{\textbf{Memory Layout and Interleaving:}}
While high- performance memory systems rely on channel interleaving to maximize the memory bandwidth, \textit{DRIM} adopts a different approach through maximizing spatial locality and allocating memory as close to their corresponding operands as possible. The main goal is to reduce the data movement across memory modules and hence reducing operations latency and energy costs. As exposing a programmer directly to the layout of memory is challenging, \textit{DRIM} architecture can rely on compiler passes that take memory layout and the program as input, then assign physical addresses that are adequate to each operation without impacting the symantics of the application.

$\bullet$\underline{\textbf{Reliability:}}
Many ECC-enabled DIMMs rely on calculating some hamming code at the memory controller and use it to correct any soft errors. Unfortunately, such a feature is not available for \textit{DRIM} as the data being processed are not visible to the memory controller. Note that this issue is common across all PIM designs. To overcome this issue, \textit{DRIM} can potentially augment each row with additional ECC bits that can be calculated and verified at the memory module level or bank level. Augmenting \textit{DRIM} with reliability guarantees is left as future work.

$\bullet$\underline{\textbf{Cache Coherence:}}
When \textit{DRIM} updates data directly in memory, there could be stale copies of the updated memory locations in the cache, thus data inconsistency issues may arise. Similarly, if the processor updates cached copies from memory locations that \textit{DRIM} will process later, \textit{DRIM} could actually use wrong/stale values. There are several ways to solve such issues in off-chip accelerators, the most common one is to rely on operating system (OS) to unmap the physical pages accessible by \textit{DRIM} from any process that can run while computing in \textit{DRIM}. \vspace{-1em}

\section{Conclusion}
In this work, we presented \textit{DRIM}, as a high-throughput and energy-efficient PIM architecture to address some of the existing issues in state-of-the-art DRAM-based acceleration solutions for performing bulk bit-wise X(N)OR-based operations i.e. limited throughput, row initialization, reliability concerns, etc. incurring less than 10\%  on top of commodity DRAM chip. 


\bibliographystyle{IEEEtran}
\bibliography{refs}

\balance

\end{document}